\documentclass[twocolumn,showpacs,preprintnumbers,amsmath,amssymb]{revtex4}

\usepackage{graphicx}
\usepackage{dcolumn}
\usepackage{bm}

\begin{document}

\title{Scale-invariance of human EEG signals in sleep}

\author{Shi-Min Cai$^{1}$}
\author{Zhao-Hui Jiang$^{1}$}
\author{Tao Zhou$^{1,2}$}
\email{zhutou@ustc.edu}
\author{Pei-Ling Zhou$^{1}$}
\author{Hui-Jie Yang$^{2}$}
\author{Bing-Hong Wang$^{2}$}

\affiliation{$^{1}$Department of Electronic Science and Technology,
University of Science and Technology of China, Hefei Anhui, 230026,
PR China\\ $^{2}$Department of Modern Physics, University of Science
and Technology of China, Hefei Anhui, 230026, PR China }

\date{\today}

\begin{abstract}
We investigate the dynamical properties of electroencephalogram
(EEG) signals of human in sleep. By using a modified random walk
method, We demonstrate that the scale-invariance is embedded in
EEG signals after a detrending procedure. Further more, we study
the dynamical evolution of probability density function (PDF) of
the detrended EEG signals by nonextensive statistical modeling. It
displays scale-independent property, which is markedly different
from the turbulent-like scale-dependent PDF evolution.
\end{abstract}


\maketitle

{\it -Introduction.} The analysis of electroencephalogram (EEG)
signals attracts extensive attentions from various research
fields, since it can not only help us to understand the dynamical
mechanism of human brain activities, but also be potentially
useful in clinics as a criterion of some neural diseases. Some
previous works have been done on human EEG signals in sleep and
other physiological states. In Refs. \cite{Fell1,Fell2,Kobayashi}
the correlation dimension and Lyapunov exponent are calculated to
characterize and discriminate the sleep stage. Lee \emph{et al.}
provides the evidence of the long-range power law correlations
embedded in EEG signals \cite{Lee}. The mean scaling exponents are
distinguished according to REM (Rapid Eye Movement), Non-REM and
awake stage, and gradually increased from stage 1 to stage 2, 3
and 4 in non-REM sleep. Hwa \emph{et al.} found the variable
scaling behavior in two regions, and described the topology plots
of scaling exponents in this two regions that reveals the spatial
structure of the nonlinear electric activity \cite{Hwa}. The
random matrix theory is performed to demonstrate the existence of
generic and subject-independent features of the ensemble of
correlation matrix extracted from human EEG signals \cite{Seba}.
Yuan \emph{et al.} found the similar long-range temporal
correlations and power-law distribution of the increment of EEG
signals after filtering out the $\alpha$ and $\beta$ wave
\cite{Yuan}. In the present paper, the Tsallis entropy is used to
analyze a series of human EEG signals in sleep.

We use the MIT-BIH polysomnography data, which is consist of
four-, six- and seven-channel polysomnographic recordings, each
with an ECG signal annotated beat-by-beat, and an EEG signal
annotated with respect to sleep stages \cite{Goldberger}. Records
have been sampled at frequency 4 kHZ. Sleep stage was annotated at
30s intervals according to the criteria of Rechtschaffen and
Kales, denoted by six discrete levels-1, 2, 3, 4 REM and awake
(stages 1, 2, 3, 4 belong to non-REM sleep) \cite{Rechtschaffen}.
In the present analysis, only the samples containing sufficient
records (at least no least than five stages) are considered. A
representative example is shown in Fig.1.

\begin{figure}
\center \scalebox{1}[0.8]{\includegraphics{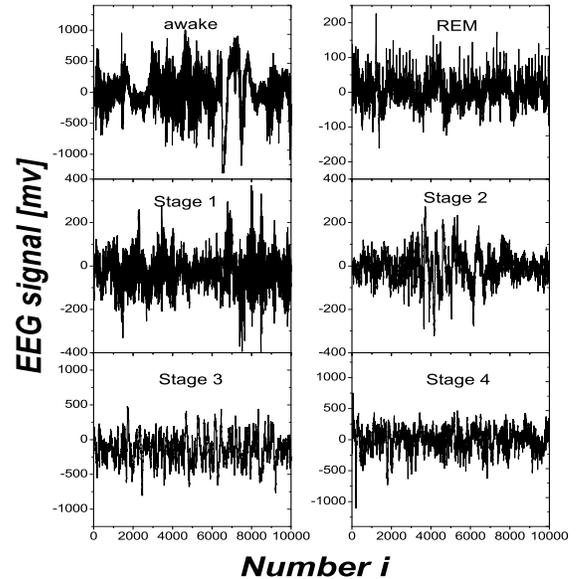}}
\caption{\label{fig:epsart}A set of representative records of EEG
signals in different stages. Each entire series includes more than
$10^{5}$ data points, while the plotted is only a small fraction.}
\end{figure}

{\it -Scale-invariance of detrended EEG signals.} Consider an EEG
series, denoted by $\{x_{i}\}$ $( i=1,2,\ldots, N)$, whose scaling
characteristics are detected through the following procedure:

\emph{Step 1:} Construct the profile series, $
Y_{j}=\sum_{i=1}^{j}x_{i}, j=1,2,\cdots, N$, and consider $Y_{j}$ as
the $``walk$ $dsiplacement"$ of a resultant random walk.

\emph{Step 2:} Divide the profile series into non-overlapping
segments with equal length and fitting each segment with a second
order polynomial function. Regard the fitting results as the trends,
a stationary series can be obtained by eliminating the trends from
the profile series.

\emph{Step 3:} After the detrending procedure, we define the
increment of this modified profile series at a scale $s$ as
$\Delta_{s}Y_j=Y^{*}_{j+s}-Y^{*}_{j}$, where $Y^{*}_{j}$ is the
deviation from the polynomial fit.

\emph{step 4:} Scale-invariance (self-similarity) in the
stationary series implies that the PDF satisfies,
$P(x,s)=\frac{1}{\sigma_{s}}P(\frac{x}{\sigma_{s}})$, where
$\sigma_{s}$ denotes the standard deviation at time scale $s$.
Obviously, $P(0,s)=P(0)\frac{1}{\sigma_s}$.

Assigning the values of parameter $s$ from $2^{1}$ to $2^{10}$,
the normalized PDFs of $\Delta_{s}Y$ exhibit scale-invariant
(self-similar) behaviors as presented in Fig. 2. That is to say,
those PDFs can be rescaled into a single master curve, as shown in
Fig. 3. The scale-invariance of the detrended EEG signals suggests
that the quasi-stationary property is embedded in the
distributions of time scales. Therefore, it helps us to search for
stable distributions to mimic them.

\begin{figure} \center
\scalebox{1}[0.8]{\includegraphics{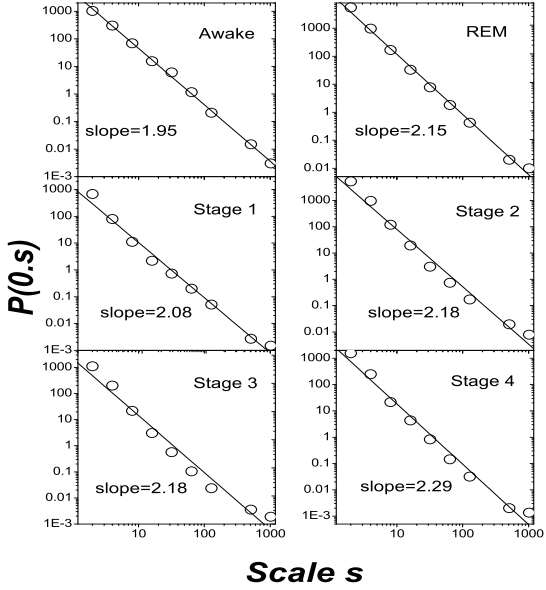}} \caption{\label{fig:epsart}
The probability of $P(0,s)$ as a function of the time sampling scale
$s$. A power-law scaling behavior is observed for about three order
of magnitude.}
\end{figure}

\begin{figure} \center
\scalebox{1}[0.8]{\includegraphics{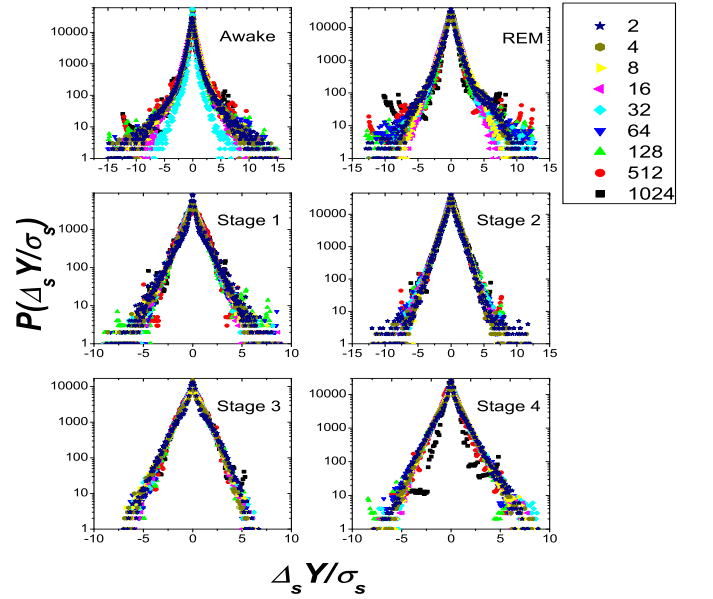}} \caption{\label{fig:epsart}
(color online) The rescaled increment's PDFs for six stages.
Obviously, curves with different time scales can well collapse into
a single master one, demonstrating the existence of quasi-stationary
property.}
\end{figure}
\begin{figure}
\center \scalebox{1}[0.8]{\includegraphics{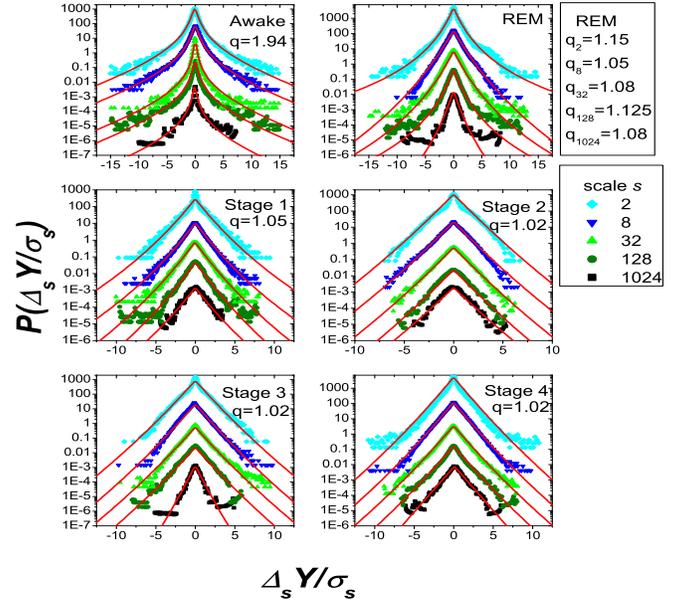}}
\caption{\label{fig:epsart} (color online) The rescaled
increment's PDFs of all stages with the approximate fit using
nonextensive statistical modeling. We use $q$-Gaussian function to
fit the awake stage, and $q$-exponential function to fit the other
five stages. The awake stage falls into l\'{e}vy regime with well
fitting parameter $q=1.94$. In the REM stage, the values of $q$ is
slight change describing as $q_{2}$, $\cdots$, $q_{1024}$, while
in the non-REM stages, parameter $q$ shows a unique value for each
specific stages.}
\end{figure}
\begin{figure}
\center \scalebox{0.8}[0.8]{\includegraphics{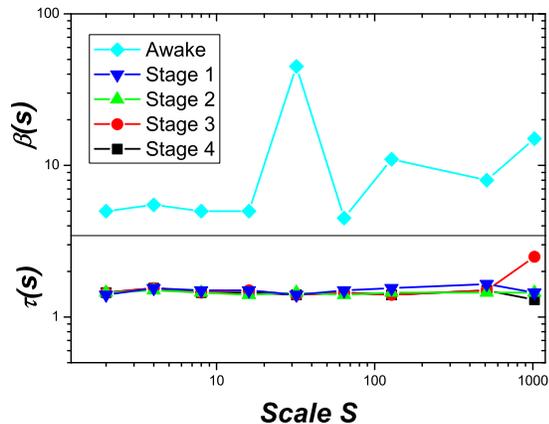}}
\caption{\label{fig:epsart} (color online) $\beta(s)$ and
$\tau(s)$ versus $s$ for awake and non-REM stages. The values of
$\beta(s)$ don't dissipate as the increasing of $s$. In
particular, $\tau(s)$ of non-REM sleep converge to an invariant
pattern.}
\end{figure}
\begin{figure}
\center \scalebox{0.7}[0.7]{\includegraphics{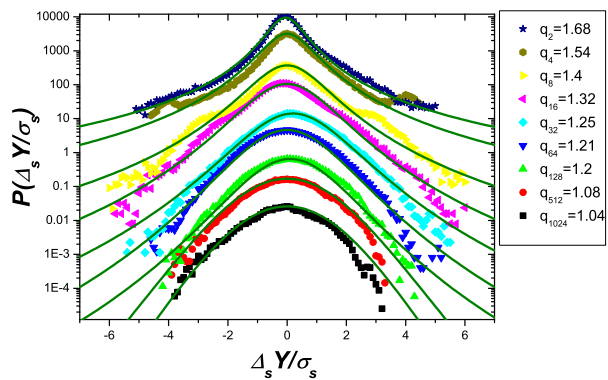}}
\caption{\label{fig:epsart} (color online) The increment's PDF of
randomized series of awake stage and fitting curves with different
parameter $q$. The parameter $q$ rapidly approaches to Gaussian
regime ($q=1$) as the time scale increases. For clarity, we
shifted the distributions through dividing them by their standard
deviation}
\end{figure}

{\it -Nonextensive statistical modeling of detrended EEG signals.}
From the results sketched in the preceding section, herein we use
the Tsallis entropy to model the PDFs. The Tsallis entropy is
introduced by Tsallis through generalizing the standard
Boltzmann-Gibbs theory \cite{Tsallis}, which reads,

\begin{equation}
S_{q}=k\frac{1-\int dx [p(x)]^{q}}{q-1}, \ \ \ \ \ (\int dx p(x)=1;
q\in R).
\end{equation}

In the limit $q\rightarrow1$, $S_{q}$ degenerates to the usual
Boltzmann-Gibbs-Shannon entropy as $S_{1}=-\int p(x)ln[p(x)]dx$. The
optimization (e. g. maximize $S_{q}$ if $q>0$, and minimize $S_{q}$
if $q<0$) of $S_{q}$ with the normalization condition $\int dx
p(x)=1$, as well as with the constrain $\langle\langle x^{2}
\rangle\rangle _{q}=\sigma^{2}$, leads to $q$-Gaussian distribution
\begin{equation}
G_{q}(x,s)=\frac{1}{Z_{q}(s)}\{1-\beta(s)[(1-q)(x-\overline{x}(s))^{2})]\}_{+}^{\frac{1}{1-q}
},\ \ \ \ \ (q<3),
\end{equation}
where $Z_{q}(s)$ is a normalization constant, $\beta(s)$ is
explicitly related to the variance of distribution, and the
subscript $``+"$ indicates that $G_{q}(x,s)$ is non-negative
\cite{Abe}. $G_{q \rightarrow 1}(x,s)$ recovers the usual Gaussian
distribution. The $q$-Gaussian PDF can describe a set of stable
distributions from Gaussian to L\'{e}vy regimes \cite{Levy} by
adjusting the value of $q$ with appropriate time-dependent
parameters $\beta(s)$ and $Z_{q}(s)$ \cite{Vignat}. The distribution
falls into L\'{e}vy regimes in the interval $\frac{5}{3}<q<3$, with
$q=5/3$ a critical value.

The results in Fig. 4 show us that the PDFs of awake stage falls
into the L\'{e}vy regime with $q$ being equal to 1.94. It exhibits
sharp kurtosis and long-tail distribution, distinguished from
those of REM and non-REM stages. It should be noted that we shift
the distributions through dividing by their standard deviation and
only plot part of them to make the figures clear. The specific
values of $\beta(s)$ for all scales are shown in the Fig. 5. It
interests us that $\beta(s)$ does not dissipate as the scale
increases unlike the case of $\beta(s)$ found in financial market
\cite{Cortines}. In other words, It demonstrates that the dynamics
evolution of EEG signals is not coincident with the diffusion
process described by Fokker-Planck equation.

Another significant equation of nonextensive statistical approach
is the $q$-exponential function, which reads
\begin{equation}
e_{q}(x,s)=\frac{1}{Z_{q}(s)}\{1-\tau(s)\left[(1-q)|x-\overline{x}(s)|\right]\}_{+}^{\frac{1}{1-q}},
\end{equation}
where the parameter $\tau(s)$ is the relaxation rate of
distribution. Clearly, in the limit $q\rightarrow1$,
$e_{1}(x,s)=\frac{1}{Z_{q}(s)}exp\{-\tau(s)|x-\overline{x}(s)|\}$.
Because the statistic distributions of detrended increment of EEG
signals in sleep stage exhibit an approximately exponential form,
we use the $q$-exponential model to quantifies them, as shown in
Fig. 4. The values of $q$ for the REM and non-REM stage are little
larger than 1. It means that the fluctuation of human brain
activity in sleep stage will converge to a normal exponential
pattern. In particular, the EEG signals exhibit $q$-invariant
pattern for different time scales in all the four stages within
non-REM sleep. The relaxation rates of distributions are also
approximately invariant, as shown in the Fig. 5. However, in the
REM stage, the values of $q$ are slight change because of fitting
the tail of distribution in different time scales, and the model
can only well fit the center distribution. It suggests that brain
electric activity in the REM stage may work in a more complex
pattern than awake and non-REM stage for the acute neural activity
\cite{Kandel}.

The nonextensive statistical approach modeling the detrended
increment's PDF of EEG signals with an invariant parameter $q$
demonstrates the scale-independent property of the system. In
order to further test the existence of this observed property, we
randomize the empirical series of awake stage by shuffling
\cite{Theiler1,Theiler2} and show a fit for this artificial
distributions at different scales in the Fig. 6. Clearly, the
parameter $q$ will approach to Gaussian regime ($q=1$) as the time
scale increases. This result strongly supports the
scale-independent property of human brain activity in sleep is
remarkably different from the turbulent-like scale-dependent
evolution \cite{Lin}.

{\it -Conclusion.} In this work, several dynamical properties of
human EEG signals in sleep are investigated. We firstly use a
modified random walk method to construct the profile series
including the information of EEG signals. After a detrending
procedure, we obtain the stationary series and define the
increments of the resultant random walk at multiple scales. In
order to characterize the dynamical process of brain electronic
activity, we then study the $P(0,s)$ of the PDF of normalized
increments as a function of $s$. With this choice we investigate
the point of each probability distribution that is least effected
by the noise introduced by experimental data set. The
scale-invariance in both awake and sleep stages are obtained, thus
one can rescale the distributions at different scales into a
single master curve.

Aim at this property, we use nonextensive statistical approach to
model these processes. The dynamical evolution of detrended
increment's PDF in awake stage can be well fitted by $q$-Gaussian
distribution with a invariant parameter $q=1.94$. It demonstrates
that the PDFs of awake stage fall into the l\'{e}vy regime.
Contrastively, $q$-exponential distribution is used to mimic PDFs
of sleep stage. In particular, for the non-REM stage, it exhibits
scale-independent distributions, while for REM stage, it suggests
a complex distributional form with the values of $q$ slightly
different.

The statistical properties of distribution of EEG signals strongly
indicate that the process of brain electric activity is remarkably
different from tubulent-like cascade evolution. In a recent work
\cite{Lin}, Lin-Hughson proposed a tubulent-like cascade model to
mimic the human heart-rate, whose validity is, now, in the face of
challenge on critical scaling-invariance found in real human
heart-rate processes \cite{Kiyono1,Kiyono2}. Although the
electrocardiograph (ECG) and electroencephalogram are different,
they have some common statistical features. We wish this work is
helpful for the in-depth understanding about the underlying
dynamical mechanism. And, like the corresponding empirical studies
on human ECG singles \cite{Kiyono1,Kiyono2}, this work could
provide some criterions for theoretical models on human EEG
signals.

These similar results are also found in the ECG signals of human.
Since the work  presents the turbulent-like cascade heart rate
model to mimic human heart rate, Kiyono \emph{et al.} found that
the human heart rate exhibits critical scaling-invariance, of
which the dynamical evolution of increment's PDF is different from
turbulent-like PDF evolution.

Further more, some very recent works \cite{Lo,Comte} pointed out
that the sleep-wake transitions exhibit a scale-invariant patterns
and embed a self-organized criticality (see also Ref. \cite{Bak}
for the concept of self-organized criticality). Thus, the
dynamical properties of human EEG signals in sleep suggest that
human brain activity in sleep may relate with a self-organized
critical system \cite{Arcangelis}. Our empirical result, in some
extent, support those conclusions.

This work is supported by the National Natural Science Foundation
of China under Grant Nos. 70571075, 70571074 and 10635040, and the
Foundation for graduate students of the USTC under Grant No.
KD2006046.

\end{document}